\documentclass{revtex4}
\usepackage{epsfig}

\begin{document}

\title{Bounds to binding energies from the concavity of thermodynamical 
functions}

\author{B. K. Jennings \\
jennings@triumf.ca, TRIUMF, Vancouver BC, V6T2A3, Canada \\
        B. R. Barrett \\
bbarrett@physics.arizona.edu, Department of Physics, \\ 
University of Arizona, Tucson, AZ 85721, USA \\
        B. G. Giraud \\
bertrand.giraud@cea.fr, Service de Physique Th\'eorique, \\
DSM, CE Saclay, F-91191 Gif/Yvette, France}

\date{\today} 

\begin{abstract}

Sequences of experimental ground-state energies are mapped onto concave 
patterns cured from convexities due to pairing  and/or shell effects. The 
same patterns, completed by a list of excitation energies, can be used to 
give numerical estimates of the grand potential $\Omega(\beta,\mu)$ for a 
mixture of nuclei at low or moderate temperatures $T=\beta^{-1}$ and at many 
chemical potentials $\mu.$ The average nucleon number 
$\langle {\bf A} \rangle(\beta,\mu)$ then becomes a continuous variable, 
allowing extrapolations towards nuclear masses closer to drip lines. We study 
the possible concavity of several thermodynamical functions, such as the free 
energy and the average energy, as functions of $\langle {\bf A} \rangle.$ 
Concavity, when present in such functions, allows trivial interpolations and 
extrapolations providing upper and lower bounds, respectively, to binding 
energies. Such bounds define an error bar for the prediction of binding 
energies. An extrapolation scheme for such concave functions is tested. We 
conclude with numerical estimates of the binding energies of a few nuclei 
closer to drip lines.

\end{abstract}

\maketitle

\section{Introduction}

The observation of a valley of stability and the search for mass formulae 
belong to the oldest subjects studied in nuclear physics. Given the
neutron and proton numbers $N$ and $Z$ as independent variables and the
corresponding atomic number, $A \equiv N+Z,$ terms such as volume energy 
$\propto (N+Z)$, surface tension $\propto (N+Z)^{\frac{2}{3}},$ Coulomb energy 
$\propto Z(Z-1)/(N+Z)^{1/3},$ symmetry energy $\propto (N-Z)^2/(N+Z),$ etc. 
flourish in the literature, and a great deal of attention has been dedicated 
to the consideration of finer corrections, such as, for instance, terms
$s(N,Z)$ and $p(N,Z)$ that account for shell and pairing effects,
respectively. This work is motivated by the observation that the dominant
terms, namely $\propto (N+Z),$ $\propto Z(Z-1)$ and $\propto (N-Z)^2$, define
a paraboloid energy surface, notoriously {\it concave}. 

Upper and lower bounds to nuclear binding energies can be deduced from such a 
concavity, provided that deviations from concavity, possibly induced by 
subdominant terms like $\propto (N+Z)^{\frac{2}{3}},$ $s(N,Z),$ $p(N,Z),$ 
etc., can be corrected. In a previous article \cite{BGJT}, we showed how 
elementary transformations of data could generate truly concave patterns. 
This was obtained by an analysis of the table of second differences between 
binding energies, then by a removal of pairing energy, and finally by 
an {\it ad hoc}, but minimal, parabolic term added to nuclear energies, if 
necessary.

Concavity is also a property of several thermodynamical functions. An 
extension of the analysis at zero temperature \cite{BGJT} to a finite 
temperature theory is in order. This extension is the main subject of the 
present paper. For the sake of simplicity this paper, like \cite{BGJT}, 
considers only sequences of isotopes and, thus, takes advantage of concavity 
with respect to $N$ only; $Z$ is frozen. A generalization to concavity with 
respect to both $N$ and $Z$ is left to future work. 

In Sec. II we briefly recall the method, explained in \cite{BGJT}, for the 
tuning of actual experimental data into concave patterns. In Sec. III we 
discuss properties of that grand potential, $\Omega(\beta,\mu),$ which can be 
deduced from the experimental data after their tuning. Other thermodynamical 
functions will also be considered, and their concavity will be tested.
Bounds will be found, and an error bar for predictions will be estimated. 
Section IV contains a brief discussion of the problems raised by 
extrapolations of concave functions. A discussion and conclusion are given 
in Sec. V.

\section{Concavity with experimental ground-state energies}

Our argument is illustrated numerically, by using a sequence of isotopic
ground-state binding energies, $-E_A.$ Consider the table from 
$^{110}$Sn to $^{137}$Sn, studied in \cite{BGJT} at zero temperature, 
because we later want to extend it at finite temperature by including 
the energies of excited states. It reads, in keV,

\noindent
\{934562,  942743,  953529,  961272,  971571,  979117,  988680,
  995625, 1004952, 1011437. 1020544, 1026715, 1035529, 1041475,
 1049962, 1055695, 1063889, 1069448, 1077348, 1082713, 1090400, 
 1095615, 1102917, 1105335, 1109075, 1111310, 1115087, 1117150\}.

\noindent
It should be noted that such data come from Refs. \cite{Wap95,Wap97,pe,Wap03}.
These sources often quote the binding energy per nucleon instead of the 
total binding energy itself and such values per nucleon are given to 
varying numbers of significant figures, from four to seven. Consequently, 
even though we quote and use all our binding energies to six or seven 
significant figures, for consistency reasons and for ease of performing our 
calculations, our values are generally accurate only to the order of tens 
of keV. It must be understood that all energies stated in this paper are in 
units of keV.

Despite a well known linear trend because of a ``not too much fluctuating 
average energy per nucleon'', this list of energies is far from making a 
smooth pattern. It is even less of a concave one. The sequence of $26$ second 
differences (SDs), $E_{A+1}-2 E_A+E_{A-1},$

\noindent
\{-2606, 3043, -2556, 2753, -2018, 2619, -2382, 2841, -2622, 2937, -2643,
   2868, -2542, 2755, -2460, 2634, -2341, 2535, -2322, 2473, -2088, 4885,
  -1323, 1506, -1542, 1714\},

\noindent
gives estimates of the ``curvatures'' of the pattern. It is far from 
containing only positive numbers. On the contrary, its signs alternate, 
systematically. The wiggling between SD's centered at odd and even nuclei
(or staggering) has, roughly speaking, a constant amplitude. (Notice, however,
the maximum SD, 4885, due to the shell closure at $^{132}$Sn, and the
weakening of the numbers beyond $^{132}$Sn.)

These alternating signs are obviously due to the gains of binding for even Sn 
nuclei because of pairing. Add to each {\it even} nucleus energy a fixed 
number, for example $p(N,Z)=1050$ keV, to suppress the increase of binding 
due to pairing. The resulting list of SDs is attenuated by an amount equal to
$\pm 2p$ and now reads,

\noindent
\{-506, 943, -456, 653, 82, 519, -282, 741, -522, 837, -543, 768, -442, 
   655, -360, 534, -241, 435, -222, 373, 12, 2785, 777, -594, 558, -386\}.

\noindent
All numbers are now significantly smaller than their partners in the previous 
list of SD's, except for the ``spike'' at $^{132}$Sn. The latter is 
positive, and, hence causes no deviation from concavity. The interesting
point is the most negative number in the list, namely $-594$ keV. All negative 
curvatures can be converted into positive ones if we add to every energy an 
artificial, parabolic correction, which was chosen in \cite{BGJT} as, 
$P \times (A-118)^2,$ with $P=300$ keV. Incidentally, the lowest point of the 
parabola is arbitrary, because SDs will increase by just a constant, namely 
twice the coefficient $P$ of the $A^2$ term. 

Hence, after such a $600$ keV shift, the previous sequence of SDs
becomes entirely positive,

\noindent
\{94, 1543, 144, 1253, 682, 1119, 318, 1341, 78, 1437, 57, 1368, 158, 1255,
  240, 1134, 359, 1035, 378, 973, 612, 3385, 1377, 6, 1158, 214\},

In short, a ``concavity ensuring'' manipulation for the Sn isotope 
energies consists in replacing each energy $E_A$ by 
$E^{\, \prime}_A=E_A + 1050 \times Mod[A+1,2] + 300 \times (A-118)^2.$ This 
indeed creates a concave pattern. The list of such tuned energies, 
$-E^{\, \prime}_A,$ reads,

\noindent
\{914312,    928043,   941679,   953772,   965721,   976417,   986430,  995325,
  1003902,  1011137,  1018294,  1024015,  1029679,  1033975,  1038112,
  1040995,  1043639,  1045148,  1046298,  1046413,  1046150,  1044915,
  1043067,  1037835,  1031225,  1024610,  1016837,  1008850\}.

\noindent
The choice of the two parameters, p=1050 keV and P=300 keV, is empirical: one 
must find a pairing correction leading to a minimal parabolic correction 
inducing concavity. Other choices for $\{p,P\}$ are obviously possible. 
Anyhow, such parameters must be readjusted for different regions of the table 
of nuclei.

When concavity is obtained, it is trivial to see that extrapolations from 
two points on the concave pattern allow predictions of lower bounds to nuclear 
energies. In the same way, it is trivial that interpolations provide upper 
bounds. Then, from such bounds for energies $E^{\, \prime},$ one recovers 
bounds, of strictly the same quality, for the physical energies $E.$ This 
is obtained by subtracting from $E'$ bounds their ``tuning terms''. 

In \cite{BGJT} we showed that the quality of such bounds is indeed good, and
that interpolations and extrapolations from the raw, non concave pattern, are 
less satisfactory. But there is a more profound reason why a concave pattern 
is necessary. Indeed, several thermodynamical functions, governed by theorems 
proving their concavity, have a notoriously singular limit at zero 
temperature: they become non analytical and are just piecewise continuous. 
Their limit plots are made of segments; derivatives are discontinuous at 
turning points. Because of the staggering effect, the {\it concave envelope}
of the raw pattern of $E_A$ would contain only the even isotopes. Concavity 
is thus necessary for a theory which must accommodate both odd and even 
nuclei. The thermodynamical functions studied in the next section, Sec. III, 
therefore, preferably use concave energies $E^{\, \prime}_A$ and the 
corresponding excited state energies $E^{\, \prime}_{nA}.$

\section{Concavity with thermodynamical functions}

Consider the particle number operator ${\bf A}$ and a familiar nuclear 
Hamiltonian ${\bf H}=\sum_{i=1}^A t_i + \sum_{i>j=1}^A v_{ij},$ where $A,$ 
$t$ and $v$ are the mass number, one-body kinetic energy and two-body 
interaction, respectively. Nuclear data tables give precise values for 
a large number of lowest lying eigenvalues $E_{nA}$ of ${\bf H},$
for many nuclei. One may thus reasonably estimate the grand partition function,
\begin{equation}
{\cal Z}(\mu,\beta) = {\rm Tr}\, 
\exp\left[ \beta \left( \mu {\bf A} - {\bf H} \right)  \right] =
\sum_{nA} (2 j_{nA}+1) 
\exp\left[ \beta \left( \mu A - E_{nA} \right)  \right],
\label{definZ}
\end{equation} 
provided that 
i) the temperature, $T=\beta^{-1},$ is low enough to allow a truncation
of the spectrum to include only those states provided by the tables and 
ii) the chemical potential, $\mu,$ selects mainly those nuclei in which we
are interested. Let $\langle \ \rangle$ denote, as usual, a statistical
average. The (equilibrium!) density operator in Fock space,
$\rho={\cal Z}^{-1} \exp\left[ \beta\, (\mu {\bf A} - {\bf H}) \right],$
ensures that the following grand potential,  
$\Omega(\mu,\beta) = \langle\, ( \mu {\bf A} - {\bf H} )\, \rangle + T\, S,$
is maximum in the space of many-body density matrices 
with unit trace, since, by definition,
$\langle {\bf A} \rangle = {\rm Tr}\, \rho\, {\bf A},$ 
$\langle {\bf H} \rangle = {\rm Tr}\, \rho\, {\bf H},$ with the entropy,
$S=                       -{\rm Tr}\, (\rho\, \log \rho).$
Our use of a slightly non-traditional $\Omega$ makes the upcoming proofs of 
concavity somewhat easier. This grand potential also reads,
\begin{equation}
\Omega(\mu,\beta) = \beta^{-1} \ln\, {\cal Z} = 
\beta^{-1} \ln \left\{ \sum_{nA} (2 j_{nA}+1) 
\exp\left[ \beta \left( \mu A - E_{nA} \right)  \right] \right\},
\end{equation}

Trivial manipulations then give the relevant statistical averages 
$\langle \ \rangle$ of particle numbers and energies, together with their 
derivatives and fluctuations, 
\begin{equation}
\partial \Omega / \partial \mu = \langle {\bf A} \rangle =  \sum_A A\ p_A,
\ \ \ p_A= {\cal Z}^{-1} \sum_n (2 j_{nA}+1) 
\exp\left[ \beta \left( \mu A - E_{nA} \right)  \right], 
\end{equation}
and
\begin{equation}
\partial^2 \Omega / \partial \mu^2 =
\partial \langle {\bf A} \rangle / \partial \mu = \beta\, 
\left(\, \langle {\bf A}^2 \rangle - \langle {\bf A} \rangle^2\, \right),
\label{flucA}
\end{equation}
then
\begin{equation}
\partial \Omega / \partial T = S = \ln {\cal Z} - 
\beta\, \langle\, (\mu {\bf A} - {\bf H})\, \rangle,
\end{equation}
or as well,
\begin{equation}
\langle {\bf H} \rangle = {\cal Z}^{-1} \sum_{nA} (2 j_{nA}+1) 
E_{nA}\, \exp\left[ \beta \left(\mu A - E_{nA} \right)  \right].
\end{equation}
Furthermore,
\begin{equation}
\partial^2 \Omega / \partial T^2 = \beta^3 \left[
\langle\, \left( \mu {\bf A} - {\bf H} \right)^2 \, \rangle -
\langle\, \left( \mu {\bf A} - {\bf H} \right)   \, \rangle^2 \right],
\label{flucB} 
\end{equation}
and
\begin{equation}
\partial^2 \Omega / (\partial \mu\, \partial T) = -\beta^2 \left[
\langle\, {\bf A} \left( \mu {\bf A} - {\bf H} \right) \, \rangle -
\langle {\bf A} \rangle \,
\langle\, \left( \mu {\bf A} - {\bf H} \right)   \, \rangle \right].
\label{crossAB} 
\end{equation}
Since the values of $A$ are integers, all such functions ${\cal Z}, 
\Omega,...,\langle {\bf H} \rangle,...$ are clearly periodic functions of
$\mu,$ with a purely imaginary period, $2\pi i/\beta.$ This is of interest for
the study of holomorphy domains with respect to both $\mu$ and $\beta,$ but
we will freeze $\beta$ as real in the following and consider functions
of a real $\mu.$

Consider the (symmetric!) matrix of second derivatives, 
\begin{equation}
 {\cal D}= \left[ \matrix{
 \partial^2 \Omega / \partial \mu^2 & 
 \partial^2 \Omega / (\partial \mu\, \partial T) \cr
 \partial^2 \Omega / (\partial T\, \partial \mu) & 
 \partial^2 \Omega / \partial T^2 } \right].
\end{equation}
Its trace is obviously positive. Now, with short notations, 
${\bf B},$ $\Delta {\bf A}$ and  $\Delta {\bf B}$ for 
$ \mu {\bf A} - {\bf H} ,$ 
$ {\bf A} - \langle {\bf A} \rangle $ and
$ {\bf B} - \langle {\bf B} \rangle ,$ respectively, its determinant reads,
\begin{equation}
\det {\cal D}= \beta^4\, \det\left[ \matrix{
  \langle\, ( \Delta {\bf A} )^2\,                     \rangle & 
- \langle\, ( \Delta {\bf A} )\,  ( \Delta {\bf B} )\, \rangle \cr
- \langle\, ( \Delta {\bf A} )\,  ( \Delta {\bf B} )\, \rangle & 
  \langle\, ( \Delta {\bf B} )^2\,                     \rangle } \right].
\end{equation}
A trivial use of Schwarz's inequality shows that $\det {\cal D}$ is positive.
Thus, ${\cal D}$ is a positive definite matrix. In other terms, $\Omega$ is a 
concave function of $\mu$ and $T.$ In turn, the double Legendre transform,
with respect to both $\mu$ and $T,$
\begin{equation}
\mu\ \partial \Omega / \partial \mu + T\, \partial \Omega / \partial T - 
\Omega = \langle {\bf H} \rangle,
\end{equation}
shows that $\langle {\bf H} \rangle$ is a concave function of both 
$\langle {\bf A} \rangle$ and $S,$ the conjugate variables of $\mu$ and $T,$
respectively.

In the following, we do not perform the full, double Legendre transform. We 
rather retain an intermediate representation, with $\langle {\bf A} \rangle$ 
and either $T$ or $\beta.$ We stay with real variables and functions. We 
can stress that, while ${\bf A}$ has a discrete spectrum, conversely
$\langle {\bf A} \rangle$ is continuous, a monotonically increasing function
of $\mu,$ smooth provided $\beta$ is finite. The monotonicity results from
Eq. (\ref{flucA}). Actually, at low temperatures, strong derivatives signal
the onset of discrete jumps due to the integer spectrum of ${\bf A},$ but we
may stay away from this ``jumpy'' regime in the following, temporarily at
least. Anyhow, at any fixed, finite $\beta,$ the smoothness and monotonicity
of $\langle {\bf A} \rangle$ with respect to $\mu$ allows a reasonably easy
numerical calculation of the inverse function
$\mu\left( \langle {\bf A} \rangle \right).$ A main argument of this section
is thus, at fixed temperatures, to use $\langle {\bf A} \rangle$ as a 
continuous variable and attempt extrapolations towards unknown nuclei.

For this, given a fixed value of $T,$ we keep track of
$\langle {\bf A} \rangle$ and $\langle {\bf H} \rangle$ as functions of $\mu.$
Since the functional inversion from $\langle {\bf A} \rangle (\mu)$ to
$\mu(\langle {\bf A} \rangle)$ is reasonably easy, we plot 
$\langle {\bf H} \rangle$ in terms of $\langle {\bf A} \rangle$ and can
attempt an extrapolation for further values of $\langle {\bf A} \rangle.$
This extrapolation can be considered as a candidate for a mass formula, at
that finite temperature $T.$

According to Eq. (\ref{flucB}), the average, constrained energy,
$\langle \left({\bf H} - \mu {\bf A}\right) \rangle,$ is a monotonically
decreasing function of $\beta.$ Furthermore, for negative chemical potentials
$\mu$ at least, and more generally if $A$ has an upper bound, the operator,
${\bf H} - \mu {\bf A},$ is bounded from below. Therefore there is a
convergence of the process consisting in
i) extrapolating with respect to $\mu$ both 
$\langle \left({\bf H} - \mu {\bf A}\right) \rangle$ 
and $\langle {\bf A} \rangle$ for fixed values of $\beta,$ then
ii) eliminating $\mu$ to generate the $\beta$-parametrized ``mass formula'' 
$\langle {\bf H} \rangle \left(\langle {\bf A} \rangle, \beta \right),$ 
and finally iii) considering the limit of this mass formula when 
$\beta \rightarrow +\infty.$ Alternately, it is equivalent, and maybe more
efficient, to first eliminate $\mu$ and then extrapolate the ``mass formula''
$\langle {\bf H} \rangle \left(\langle {\bf A} \rangle, \beta \right),$ first
with respect to $\langle {\bf A} \rangle,$ then with respect to $\beta.$

Are there concavity properties in this intermediate representation? Clearly, 
a simple Legendre transform of $\Omega,$ with respect to $\mu$ only, returns 
a free energy, $\langle {\bf H} \rangle - T\, S,$ as a concave function of 
$\langle {\bf A} \rangle$ and $T.$ If $T$ is low enough to allow the
product $T\, S$ to be neglected, then, at fixed $T,$ one may accept that
$\langle {\bf H} \rangle$ is an ``almost'' concave function of 
$\langle {\bf A} \rangle.$ This assumption will be tested by the numerical 
results which follow. Incidentally, a straightforward calculation yields,
\begin{equation}
\frac{\partial^2 \langle {\bf H} \rangle} {\partial \langle {\bf A} \rangle^2}
\left( \langle {\bf A} \rangle, \beta \right)\, \propto\, 
\left(\, \langle {\bf A}^2 \rangle - \langle {\bf A}\rangle^2\, \right)\, 
\langle {\bf A}^2 {\bf H} \rangle + 
\left(\, \langle {\bf A} \rangle\, \langle {\bf A}^2 \rangle - 
\langle {\bf A}^3 \rangle\, \right)\, 
\langle {\bf A} {\bf H} \rangle + 
\left(\, \langle {\bf A} \rangle\, \langle {\bf A}^3 \rangle - 
  \langle {\bf A}^2 \rangle^2\, \right)\, 
\langle {\bf H} \rangle,
\label{secderi}
\end{equation}
with a positive factor,
$\left( \langle {\bf A}^2 \rangle - \langle {\bf A} \rangle^2 \right)^{-3}.$
This simplifies into
\begin{equation}
\frac{\partial^2 \langle {\bf H} \rangle} {\partial \langle {\bf A} \rangle^2}
\left( \langle {\bf A} \rangle, \beta \right)\, \propto\,
\langle\, (\Delta {\bf A})^2\, \rangle\ 
\langle\, (\Delta {\bf A})^2\ \Delta {\bf H}\, \rangle -
\langle\, (\Delta {\bf A})^3\, \rangle\ 
\langle\, \Delta {\bf A}\     \Delta {\bf H}\, \rangle, 
\label{secderisimpl}
\end{equation}
if one uses the already defined, centered operator 
$\Delta {\bf A}$ and the similarly centered operator 
$\Delta {\bf H}={\bf H}-\langle {\bf H} \rangle.$
From Eqs. (\ref{secderi},\ref{secderisimpl}), concavity is unclear; we shall 
have to test it numerically.

The results, which follow, use \cite{istps} and also the tables, already 
quoted \cite{Wap95,Wap97,pe,Wap03}. Using the first $10$ levels of $^{110}$Sn 
to $^{137}$Sn, successively, and fewer levels if less than 10 are known, we 
calculate ${\cal Z},$ see Eq. (\ref{definZ}). For those rare cases where the 
spin $j_{nA}$ is ambiguous we choose the lowest of the suggested spins. If the 
spin is completely unknown, we set it to be either $0$ or $1/2$, according to 
$A.$ These tactics minimize the statistical influence of such rare cases. For 
a future independent treatment of the proton and neutron numbers, we actually
use a neutron number operator ${\bf N}={\bf A}-50$ rather than ${\bf A}$ in
Eqs. (\ref{definZ})-(\ref{secderi}); our chemical potential $\mu$ is truly
for neutrons. However, for easier reading, the upcoming plots use
$\langle {\bf A} \rangle=\langle {\bf N} \rangle+50.$

The lower and upper curves in the left part of Fig. 1 are the plots of the 
function $\langle {\bf H} \rangle(\langle {\bf A} \rangle)$ when $T=20$ keV
and $T=2$ MeV, respectively. The increase of $\langle {\bf H} \rangle$ when
$T$ increases is transparent. Note, however, that a lack of excited states in
our data base beyond $^{132}$Sn weakens this temperature effect. 
Also striking is the apparent concavity of both curves. It is found that the
lower temperature, $20$ keV, is low enough to allow $\langle {\bf H} \rangle$
to run, in practice, through all the experimental energies for the
{\it even} nuclei; the function follows the {\it concave envelope} of the
experimental pattern. For graphical convenience, Fig. 1 shows the plots for
$124.7 \le \langle {\bf A} \rangle \le 135.3$ only, but the same observations
hold for full plots, with $110 \le \langle {\bf A} \rangle \le 137.$ As a
test, we also calculated $\langle {\bf H} \rangle$ when the levels of
$^{132}$Sn are omitted from the trace sum, Eq. (\ref{definZ}), see the right
part of Fig. 1, where now the concave envelope goes through the dots
representing two odd nuclei, $^{131}$Sn and $^{133}$Sn, and ignores the dot
representing $^{132}$Sn. Similar verifications of other concave envelopes
were obtained by removing other nuclei when calculating ${\cal Z}.$

Then Fig. 2 shows what happens when one calculates a tuned partition
function, ${\cal Z}',$ and all resulting functions $S',$ 
$\langle {\bf A} \rangle',$ $\langle {\bf H} \rangle',$ etc. by using concave
tuned energies, 
$E'_{nA}=E_{nA}+ 1050 \times  Mod[A+1,2] + 300 \times (A-118)^2.$ The
dots in Fig. 2 now represent ground-state tuned energies; hence, concave
envelopes do not eliminate odd nuclei. Now both even and odd nuclei can be
accounted for by the function $\langle {\bf H} \rangle',$
except, as shown by the right part of Fig. 2, for those nuclei which have
been voluntarily omitted from the trace, Eq. (\ref{definZ}).

To verify whether concave envelopes and concavity result from negligible
values of the entropy term in the free energy, or, more precisely, 
negligible values of its second derivative, a calculation of $T\, S,$ for
bare data, and of its partner $T\, S',$ for tuned data, is in order.
Figure 3 shows decimal log plots of the ratio, 
$|\, T\, S'/\langle {\bf H} \rangle'\, |,$ as a function of
$\langle {\bf A} \rangle,$ for $T=100,$ $300$ and $2000$ keV, respectively.
At the higher temperature, $2$ MeV, an approximately constant ratio is
observed, except for edge effects at both ends of our data base. The almost
constant ratio favors concavity. Moreover, the ratio typically does not
exceed $10^{-2}$. This is small, but might not be small enough, because then
$T\, S^{\, \prime}$ remains close to $10$ MeV, the order of magnitude of
$|E_{A+1}-E_A|,$ a first difference and a binding energy per nucleon. For
$T=300$ keV, the ratio becomes smaller, $\sim 10^{-3}$ or less, but it
acquires some structure, because of the low level density of $^{132}$Sn in
particular. Again lower ratios, smaller than $4 \times 10^{-4},$ are found for
$T=100$ keV, but now with strong variations. It is clear that odd nuclei,
because of their higher level density due to the absence of a  pairing gap in
their spectra, induce maxima. As a consequence of such strong variations, 
translating into strong first and second derivatives, it is not excluded that
SDs coming from $T\, S^{\, \prime}$ might prevent 
$\langle {\bf H} \rangle'$ from having the concavity property of the free 
energy, $\langle {\bf H} \rangle' - T\, S^{\, \prime}.$ Obviously, the same 
difficulty might arise with $\langle {\bf H} \rangle$ and its concave 
envelopes. We must therefore numerically calculate second derivatives, see 
Eqs. (\ref{secderi},\ref{secderisimpl}).

The left part of Fig. 4 shows plots of the right-hand side
of Eqs. (\ref{secderi},\ref{secderisimpl}), multiplied by 
$\left( \langle {\bf A}^2 \rangle' - \langle {\bf A} \rangle'^2 \right)^{-2}.$
This is for tuned data. The additional denominator, 
$\left( \langle {\bf A}^2 \rangle' - \langle {\bf A} \rangle'^2 \right),$ is 
omitted for graphical reasons. The upper full line represents the situation
when $T=1$ MeV; the second derivative,
$\partial^2 \langle {\bf H} \rangle' / \partial \langle {\bf A} \rangle'^2,$
remains positive and does so until $T \sim 750$ keV, see the dashed line in
the left-hand side of the figure. For lower temperatures, however, negative
values appear. For instance, the lower full curve, corresponding to
$T=500$ keV, indicates small, but definitely negative values between
$^{130}$Sn and $^{131}$Sn. Our numerical tests show that the occurrence of 
such negative, actually moderate, values for lower temperatures
seems to be frequent, while not systematic. Furthermore, such ``negativity 
accidents'' turn out to be worse if we use untuned data, for the obvious 
reason that the untuned data lacked concavity in the first place.

A possible reason for the negativity accidents with tuned data might be that 
the fluctuation of ${\bf A}$ is not large enough to justify our use of 
$\langle {\bf A} \rangle'$ as a continuous variable. Since it interpolates 
between integers, a fluctuation of order $\sim 1,$ or at least $\sim .5,$ 
might be necessary. As shown by the three plots in the right part of Fig. 4,
corresponding to $T=1000,$ $500$ and $100$ keV from the upper to the
lower plots, respectively, a minimum temperature is needed to avoid too small 
a fluctuation of the particle number. Furthermore, the low level density in 
$^{132}$Sn obviously reinforces a reduction of fluctuations and the small SDs 
obtained by us in the tuned pattern of ground-state energies allow localized 
deviations of concavity. The deviations are too weak to appear on Fig. 2, 
however.

At this stage, the situation can be summarized as follows. On the one hand, 
the tuned pattern of experimental energies shows concavity, but the concavity
of $\langle {\bf H} \rangle'$ as a function of $\langle {\bf A} \rangle'$ is 
not sure, although it seems to occur most of the time. On the other hand, we 
have a theorem proving concavity for the free energy, 
either ${\bf F} \equiv \langle {\bf H} \rangle - T\, S = \mu\, 
\langle {\bf A} \rangle - \Omega $
or ${\bf F}' \equiv \langle {\bf H} \rangle' - T\, S' = \mu\, 
\langle {\bf A} \rangle' - \Omega', $ as functions of $T$ and 
$\langle {\bf A} \rangle$ or $\langle {\bf A} \rangle',$ respectively.
For instance, elementary derivations show that, in that representation where
$\langle {\bf A} \rangle$ (or $\langle {\bf A} \rangle'$) and $\beta$ are the 
primary variables, 
\begin{equation}
\frac{\partial{\bf F}'}{\partial \beta} = \beta^{-2}\, S',\ \ \ \  
\frac{\partial{\bf F}'}{\partial \langle {\bf A} \rangle'} = \mu\ \ \ \  
{\rm and}\ \ \ \ 
\frac{\partial^2{\bf F}'}{\partial \langle {\bf A} \rangle'^2} = \frac{T}
{ \langle {\bf A}^2 \rangle' - \langle {\bf A} \rangle'^2 }\, .
\end{equation}
The removal of the entropy term, leading from the free energy to the 
energy, can destroy the concavity, somewhat weakly. 

We are not much interested in ${\bf F},$ because, as stated earlier, its 
concavity skips odd nuclei. Thus, the remainder of this section will consider 
${\bf F}',$ at fixed $T,$ and we shall assume that $T$ is low enough to make 
$T S'$ small with respect to $\langle {\bf H} \rangle'.$ This approximation 
can be verified later, in due time. Our rationale will be that 
$\langle {\bf H} \rangle',$ even though it might deviate from concavity, will 
stay close enough to the concave ${\bf F}'.$ Their difference, $T S',$ 
a positive quantity, will define an error bar between a lower bound ${\bf F}'$ 
and an upper bound $\langle {\bf H} \rangle'$ for ground-state energies.

Consider Fig. 5. The dots represent the ground-state tuned energies,
the upper full curve is the plot of 
$\langle {\bf H} \rangle'(\langle {\bf A} \rangle')$ when $T=1$ MeV, and the 
lower dashed curve is the plot of ${\bf F}'(\langle {\bf A} \rangle')$ at the 
same temperature. There is no need to stress that two such curves make a band
defining upper and lower bounds for the experimental energies. Moreover, a 
similar, but narrower band is obtained if $T$ decreases. The lower full curve 
and the upper dashed one in Fig. 5 represent $\langle {\bf H} \rangle'$
and ${\bf F}',$ respectively, for $T=250$ keV. The following properties, i) 
the average energy and the free energy are increasing and decreasing, 
respectively, functions of $T,$ ii) the energy is larger than the free 
energy and iii) the entropy term by which they differ vanishes when $T$ 
vanishes, are not big surprises. It can be concluded that, in so far as 
thermodynamical functions can be calculated at low enough temperatures,
precise ``accuracy bands'' may be available. Their, hopefully analytic,
continuation for higher and/or lower values of $A$ than those available in 
nuclear data tables provides a scheme making predictions for exotic nuclei.

An estimate of the entropy term is now useful. Given $\mu$ and a {\it large}
$\beta,$ let $A_0$ and $E'_0$ correspond to that nucleus whose ground-state 
energy maximizes the exponential, $\exp[\beta\, (\mu A - E'_{0A})].$
Consider now the first subdominant exponential. It might be generated by the 
first excited state of the same nucleus, or by the ground-state of one of 
its neighbors. Let $A_1$ and $E'_1$ be its parameters and define 
$\Delta=\mu\, (A_1-A_0)-(E'_1-E'_0).$ Concavity guarantees that $\Delta < 0.$
Whenever $\beta$ is large enough, it is trivial to reduce the grand canonical
ensemble to a two state ensemble, and the entropy then boils down to 
$s=- e^{\beta\, \Delta}\, \beta\, \Delta.$ Hence, the product,
$T\, s =- e^{\beta\, \Delta}\, \Delta,$ vanishes exponentially fast when 
$\beta \rightarrow \infty.$ The rate of decrease is governed by that scale 
defined by $\Delta,$ to be extracted from the tuned data. Then one can 
estimate an order of magnitude for the difference between the free energy
and the energy. This estimate can be viewed as an error bar for the 
prediction of exotic nuclei via the present ``concavity method''.

\section{Further verification of concavity properties and, then, 
extrapolations}

After our previous illustration with the Sn isotopes, we now calculate 
thermodynamical functions from a sequence of Sm isotopes, i.e. $^{135}$Sm to 
$^{144}$Sm. We want to again observe the behavior of such thermodynamical 
functions and, furthermore, extrapolate them. The extrapolation might provide 
``predictions'' of the binding energies of lighter isotopes, from $^{128}$Sm 
to $^{134}$Sm for instance, which are quoted by the data tables 
\cite{Wap95}-\cite{Wap03}. According to the same data tables, indeed, the 
available accuracies for such lighter Sm nuclei still leave room for a large 
amount of theory.

Our list of raw data, $-E_A,$ is, from $^{128}$Sm to $^{144}$Sm,

\noindent
 \{1023616, 1034967, 1048320, 1059004, 1072104,
   1082088, 1094512, 1103976, 1116005, 1125291,
   1136834, 1145788, 1156935, 1165489, 1176615,
   1185216, 1195736\}.

\noindent
We will not use the first seven nuclei in our calculations of 
$\langle{\bf A} \rangle',$ $\langle{\bf H} \rangle',$ ${\bf F}',$ etc., 
because of limited and/or poorly known data for those nuclei. The last ten 
data define eight SD's,

\noindent
\{2743, -2257, 2589, -2193, 2593, -2572, 2525, -1919\}.

\noindent
With a pairing correction parameter, $p=1250,$ and a parabolic correction 
parameter, $P=36,$ the SD's become

\noindent
\{315, 315, 161, 379, 165, 0, 97, 653\}.

\noindent
Finally our list of tuned energies, 
\begin{equation}
E^{\, \prime}_A = E_A + 10000 \times A - 245000 + 1250 \times Mod[A+1,2] +
36 \times (A-139)^2,
\label{tuneSm}
\end{equation}
 reads, \{1600, 569, -147, -548, -788, -649, -345, -41, 360, 1414\}.

\noindent
The constant subtraction term, $-245000,$ and the linear additive one, 
$10000 \times A,$ are here just for graphical and numerical convenience; they 
do not change the SD's. The tuning process is illustrated by Fig. 6. In its 
left part, note the almost concave pattern made by the dots, the result of the 
pairing correction.

\noindent
In the right part of Fig. 6, the dots represent tuned energies, see 
Eq. (\ref{tuneSm}), from $^{132}$Sm to $^{146}$Sm, while the curve represents
a least square fit, between only $^{135}$Sm to $^{144}$Sm, by a quadratic 
polynomial. The list of errors, $E^{\, \prime}_{lsf}(A)-E'_A,$ generated by 
this most simple fit from $^{135}$Sm to $^{144}$Sm, is, 
\{-109, 46, 101, 58, 71, -79, -178, -61, 176, -24\}.
The right part of Fig. 6 shows that an extrapolation towards lighter nuclei 
can be considered. But an extrapolation towards heavier ones seems to be
much less valid, with bad lower bounds.

The corresponding energies $E^{\, \prime}_{nA}$ of excited states 
\cite{istps}, from $^{135}$Sm to $^{144}$Sm, are now used for the calculation 
of thermodynamical functions, $\Omega',$ $\langle{\bf A} \rangle',$ 
$\langle{\bf H} \rangle',$ etc. We take into account the lowest ten states 
per nucleus, whenever possible, and ambiguous spins are set to their minimum 
possible values. The result is shown in Fig. 7. It confirms all the analysis 
done with the Sn isotopes. The left part of Fig. 7 shows the fluctuations of 
$\langle{\bf A} \rangle'$ for $T=200$ keV, (upper curve), and $T=100$ keV, 
(lower curve), respectively. Except at both ends of the list of data used for 
the calculation, a fluctuation of at least $\sim .5$ is observed. This seems 
compatible with an extrapolation. Note, however, that difficulties are not 
excluded, because, indeed, the complete vanishing of the fluctuation at 
$^{135}$Sm and $^{144}$Sm is not a good omen.

Such end effects seem to be weaker in the right part of Fig. 7, which shows, 
again for the same temperatures, the behaviors of the average energy and the 
free energy. Dashed curves correspond to $T=200$ keV and full curves to 
$100$ keV, respectively. The curves above the dots represent 
$\langle{\bf H} \rangle'$ and those below them represent ${\bf F}'.$ We, thus, 
see again ``error bands'', which shrink when $T$ diminishes.

It is tempting to use the pair of curves, $\langle{\bf H} \rangle',$ ${\bf F}'$
at $100$ keV to attempt extrapolations, but the temperature is clearly too 
low; indeed, the curve for $\langle {\bf H} \rangle'$ shows the onset of its 
angular limit at $T=0,$ and a close inspection of the curve for ${\bf F}'$ at 
the same temperature, $T=100$ keV, detects ``angular trends'' as well. Hence, 
we shall use $T=150$ keV for the calculation of ${\bf F}',$ and $T=300$ keV 
for $\langle {\bf H} \rangle',$ as a compromise to obtain much less 
angular curves. As an additional precaution, we shall now use only the eight 
nuclei from $^{136}$Sm to $^{143}$Sm for the calculation of thermodynamical 
functions; we remove $^{135}$Sm because of its lack of known excited states 
and $^{144}$Sm because of shell closure effects. The difference, 
$\langle {\bf H} \rangle'_{300}-{\bf F}'_{150},$ is then plotted in Fig. 8.

The width of the band defined by $\langle {\bf H} \rangle'_{300}$  and 
${\bf F}'_{150}$ as functions of $\langle {\bf A} \rangle'$ can be estimated 
from Fig. 8 to be $\sim 800$ keV. Smaller values at both ends are obviously 
misleading, because they represent the edge effects of the truncation of the 
data base, where the chemical potential is large with both signs.

Predictions might, thus, be listed with a $\pm 400$ keV error. For a somewhat
crude analytical continuation below $^{136}$Sm, we make a least square fit by 
a cubic polynomial between $^{137}$Sm and $^{142}$Sm. Voluntarily, the fit 
does not take into account the functions $\langle {\bf H} \rangle'_{300}$  and 
${\bf F}'_{150}$ in the interval between $^{136}$Sm and $^{137}$Sm and in that 
between $^{142}$Sm and $^{143}$Sm, because we want to avoid the edge effects 
observed in Fig. 8. We accept an order $3$ for the polynomials rather than the 
simpler order, $2,$ because we want to take into account a skewness of 
the plots of $\langle {\bf H} \rangle'$ and ${\bf F}'.$ We verified that, 
despite such an odd order, such fitted polynomials still induce concavity in a
large enough interval. The polynomial fitting 
$\langle {\bf H} \rangle'_{300}$ reads,
\begin{equation}
{\cal P}_H(\langle {\bf A} \rangle') \simeq
-8.0\ (\langle {\bf A} \rangle'-150.1)\ (\langle {\bf A} \rangle'-141.1)\
(\langle {\bf A} \rangle'-137.4),
\end{equation}
and that fitting  ${\bf F}'_{150}$ reads,
\begin{equation}
{\cal P}_F(\langle {\bf A} \rangle') \simeq
-7.0\ (\langle {\bf A}\rangle'-153.1)\ (\langle {\bf A}\rangle'-142.9)\ 
(\langle {\bf A}\rangle'-136.2).
\end{equation}
The results are shown in Fig. 9, where the full curves 
represent the thermodynamical functions, as estimated numerically, and the
dashed curves represent their polynomial fits, extrapolated.

It is seen that such polynomial approximations for $\langle {\bf H} \rangle'$ 
and ${\bf F}'$ define a ``crescent'' rather than a band, see Fig. 9. The 
method, therefore, has obviously a limited validity domain, so long as a 
better extrapolation scheme is not found. As seen from Fig. 9, ``predictions''
for $^{135}$Sm and $^{134}$Sm will be reasonable, but it seems more risky to 
use this method below $^{134}$Sm. It is clear that the difficulty is due to 
the absence of known excited states for isotopes lighter than $^{136}$Sm, 
making it impossible to estimate the thermodynamical functions below 
$^{136}$Sm accurately enough.

Many prediction schemes can be provided by this method. For instance, the 
formula,
\begin{equation}
E_{pred}(A)=\frac{1}{2} \left[{\cal P}_H(A)+{\cal P}_F(A)\right]
- 10000 \times A + 245000 - 1250 \times Mod[A+1,2] - 36 \times (A-139)^2,
\label{predmid}
\end{equation}
takes the middle between the presumed upper bound, ${\cal P}_H(A),$ and the 
presumed lower bound, ${\cal P}_F(A);$ simultaneously, and necessarily, it
inverts Eq. (\ref{tuneSm}). The results for $^{133}$Sm, $^{134}$Sm, $^{135}$Sm,
and $^{144}$Sm read, respectively, 
$\{-1081622, -1094280, -1104086, -1196422\},$ to be compared with the values
taken from the mass tables, $E_A=\{-1082088, -1094512, -1103976, -1195736\}.$ 
Hence the corresponding errors are, $E_{pred}(A)-E_A=\{466, 232, -110,-686\},$
to be compared with those from $^{137}$Sm to $^{142}$Sm, 
$\{-31, -8, 98, 25, -59, -35\}$ and those, $\{-79,-44\},$ for the two nuclei, 
$^{136}$Sm and $^{143}$Sm, that were excluded from our least square fit but 
contributed to the thermodynamical estimates. As expected from Fig. 9, such 
results for $^{134}$Sm and $^{135}$Sm are quite satisfactory; they lie very 
much inside the estimated error bar, $\sim 400$ keV. Somewhat unexpectedly, a 
tolerable result is found for $^{133}$Sm. But, again as expected, if only from 
Fig. 9, a large error is found for $^{144}$Sm.

Another formula,
\begin{equation}
E_{low}(A)={\cal P}_F(A)
- 10000 \times A + 245000 - 1250 \times Mod[A+1,2] - 36 \times (A-139)^2,
\label{predF}
\end{equation}
obviously gives candidate lower bounds. For $^{133}$Sm, $^{134}$Sm and 
$^{135}$Sm, it generates the following errors, 
$E_{low}(A)-E_A=\{230, -38, -415\},$ with a serious failure for $^{133}$Sm.
But, if one does not trust such a formula beyond, say, one unit of mass below 
$^{135}$Sm, another formula,
\begin{equation}
E_{tang}(A)=
{\cal P}_F(134) + (A-134) \times \frac{d{\cal P}_F}{dA}\Big|_{A=134}
- 10000 \times A + 245000 - 1250 \times Mod[A+1,2] - 36 \times (A-139)^2,
\label{predtgF}
\end{equation}
uses concavity and the derivative, $d{\cal P}_F/dA |_{A=134} \simeq - 1619,$ 
to predict lower bounds. With ${\cal P}_F(134) \simeq 2600,$ the errors for 
$^{133}$Sm, $^{134}$Sm and $^{135}$Sm become, $\{11,-38,-620\}.$ As should be,
the first and third numbers are smaller than those obtained from 
Eq. (\ref{predF}) and the second number is unchanged. More important, the first
number, with still the wrong sign for an expected lower bound, is now so small
that the ``prediction'' is excellent if it is not a fortunate coincidence. It 
should be kept in mind that our experimental data are anyhow accurate only 
within tens of keV.

In turn, the formula,
\begin{equation}
E_{up}(A)={\cal P}_H(A)
- 10000 \times A + 245000 - 1250 \times Mod[A+1,2] - 36 \times (A-139)^2,
\label{predH}
\end{equation}
is assumed to yield upper bounds. Then concavity states that the further
formula,
\begin{equation}
E_{high}(A)=\frac{1}{2} \left[{\cal P}_H(A-1) + {\cal P}_H(A+1)\right]
- 10000 \times A + 245000 - 1250 \times Mod[A+1,2] - 36 \times (A-139)^2,
\label{secH}
\end{equation}
yields higher upper bounds. For $^{133}$Sm, $^{134}$Sm and 
$^{135}$Sm again, the respective errors read $\{702, 502, 195\}$ and 
$\{939, 715, 384\}.$ The second set shows, as expected, larger errors
than the first one.

This section, which uses Sm isotope data, has thus completely confirmed the 
theorems and properties, tested in the previous section with Sn isotope data.
A limit to the validity of the method has been found, however: experimental 
data for $^{133}$Sm, $^{134}$Sm and $^{135}$Sm can be confirmed by this 
theory, but, because of a very severe lack of excited state data below 
$^{136}$Sm, we find it unreasonable to extend the present ``thermodynamical'' 
theory all the way to $^{128}$Sm, and even less reasonable to use it for 
predicting Sm isotopes lighter than $^{128}$Sm. This does not prevent us, as 
an extension of the right part of Fig. 6, from attempting a polynomial fit for 
ground-state energies from, say, $^{135}$Sm to $^{128}$Sm and extrapolating 
for lighter nuclei. But the only ``thermodynamical'' aspect of such an 
extrapolation from just ground-state energies would be to retain the same 
error bar as that obtained from Fig. 8.

\section{Summary, discussion and conclusion}

We first recalled \cite{BGJT} how a table of ground-state energies for a 
sequence of isotopes can be converted into a concave pattern. This involved 
simple manipulations: for instance an explicit term, accounting for pairing in 
even nuclei, can be subtracted from the bindings. Similar arguments leading to 
concavity are obviously possible for isotones as well, and furthermore for 
any other sequence of neighboring nuclei in any direction across the table of 
known nuclei. Once this empirical tuning has been implemented, linear or
polynomial extra- and interpolations of the concave pattern may provide 
surprisingly accurate estimates of, or bounds for, binding energies. The terms 
which were added to induce concavity are subtracted {\it in fine}, naturally.

This work defined a more ambitious extra- and interpolation scheme, involving
thermodynamical functions from a grand canonical ensemble, because such 
functions may have rigorous concavity properties. A few theorems are available,
indeed. For instance, the free energy is a concave function of the average 
particle number. It is also a decreasing function of the temperature. We also 
found ``quasi-theorems'', more precisely strong numerical evidences, 
concerning the average energy. For instance, this average energy seems to be, 
except for minor accidents, a concave function of the average particle number.

For every given, finite temperature, we found that the average energy and the 
free energy, as functions of the average particle number, give upper and lower
bounds, respectively, for the concave envelope of the ground-state energies. 
When the temperature vanishes, both bounds converge to the exact results.

A difficulty remains for extrapolations at this vanishing temperature, however:
the analyticity of such thermodynamical functions is lost, because their limit 
is only piecewise continuous. It is, therefore, necessary to retain a minimum 
temperature if one wants to obtain practical extrapolations for the prediction 
of exotic nuclei. This is because a minimum amount of particle number 
fluctuation is obviously necessary to justify the conversion of particle 
number, an initially discrete quantity, into a continuous variable.

We, therefore, implemented numerical estimates of several thermodynamical 
functions at moderate temperatures, a few hundred keV at most. This yields a
first result, namely a ``band'', enclosing ground-state energies between
the average energy and the free energy. The width of the band defines an error
bar which can be definitely trusted when extrapolations are made.

A difficulty arises, however, because of an insufficient number of excited 
states; these, obviously, are missing at both ends of any sequence of isotopes.
More than often, only the ground-state energy is known for such neutron rich or
neutron poor nuclei. The calculation of the average and free energies is thus
possible only in an interval smaller than the interval of masses where ground 
states are known. Two tactics are then available, namely i) an extrapolation
of the sequence of ground-state energies alone and ii) an extrapolation of 
the thermodynamical functions, starting from a smaller interval.

The first approach was the subject of our previous work \cite{BGJT}, and the 
present work makes it more reliable because of the derivation of an error bar
in this paper. The present work tested the second tactic, with some success. 
But a limitation was found: because of the lack of known excited states, edge 
effects are present in the calculation of thermodynamical functions and, 
therefore, it is better to restrict polynomial fits to even a slightly smaller 
mass interval. Then, extrapolations by means of such polynomials seem to
be reliable within only two, at most three mass units towards drip lines.

It is clear that our polynomial extrapolations, while useful, do not take 
enough into account analytical properties which could be derived from the 
simplicity of an Hamiltonian of the form, 
${\bf H}=\sum_i t_i+\sum_{i>j} v_{ij},$ and further analytical properties of 
thermodynamical functions such as 
${\cal Z}={\rm Tr} \exp[\beta (\mu {\bf A} - {\bf H})],$ Eq. (\ref{definZ}).
Semi-classical approximations, for instance, are likely to improve this
method, when extrapolations are implemented. This is in our agenda.

In any case, we can make the strong conclusion that the combination of 
concavity and extrapolations of thermodynamical functions gives a systematic 
set of upper and lower bounds for the prediction of ground-state energies.

\bigskip
{\it Acknowledgments}: We thank I. Allison for helpful discussions and 
assistance with the management of the data sets.  It is a pleasure for B. R. B.
and B. G. G. to thank TRIUMF, Vancouver, B.~C., Canada, for its hospitality, 
where part of this work was done. The Natural Science and Engineering Research 
Council of Canada is thanked for financial support. TRIUMF receives federal 
funding via a contribution agreement through the National Research Council of 
Canada. B. R. B. also thanks Service de Physique Th\'eorique, Saclay, France, 
for its hospitality, where another part of this work was carried out, and the
Gesellschaft f\"ur Schwerionenforschung (GSI), Darmstadt, Germany, for its 
hospitality during the preparation of this manuscript, and acknowledges 
partial support from NSF grant PHY0555396 and from the Alexander von Humboldt 
Stiftung.

\vspace{-2cm}
\begin{figure}[htb] \centering
\mbox{  \epsfysize=120mm
         \epsffile{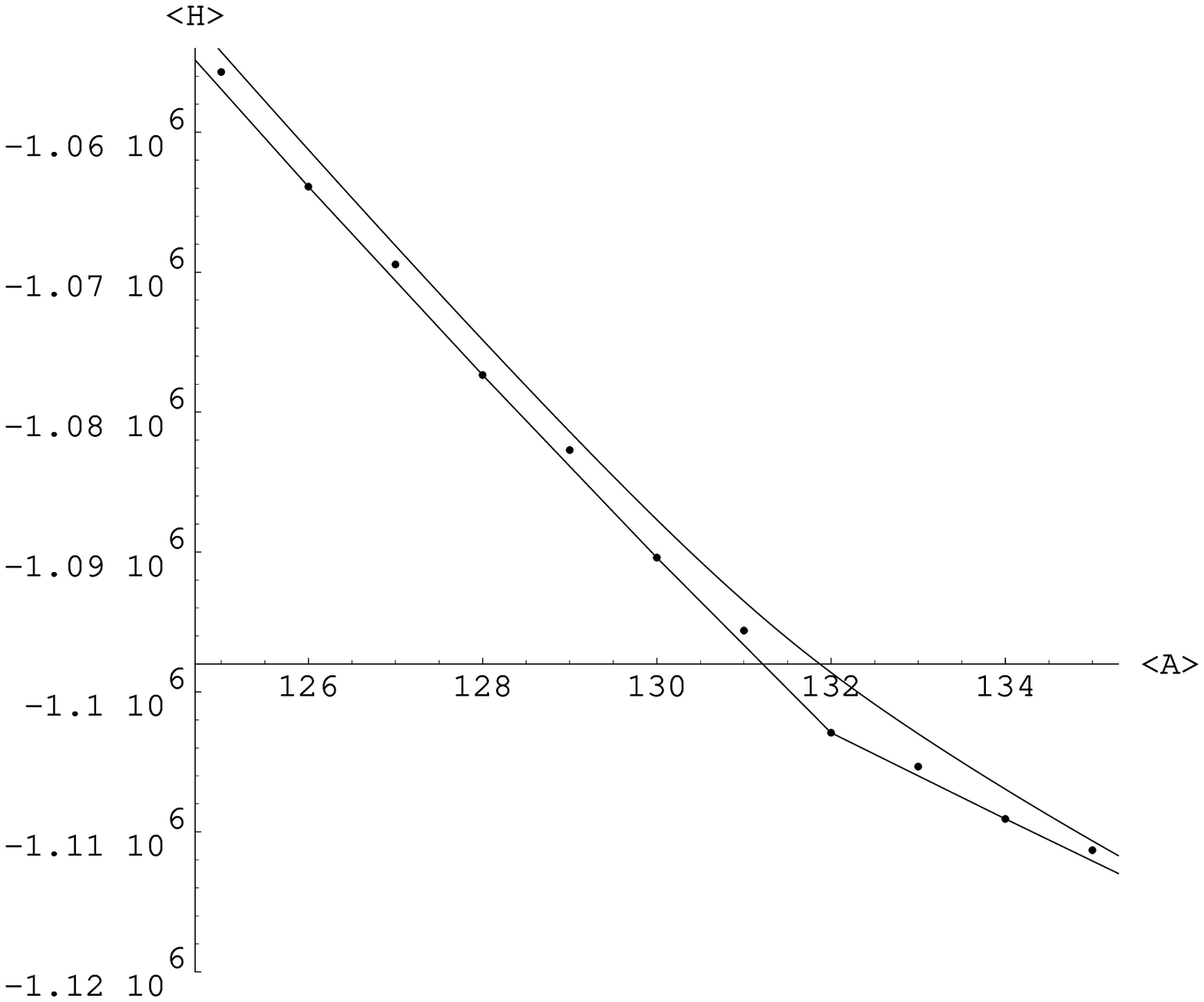}
     }
\mbox{  \epsfysize=120mm
         \epsffile{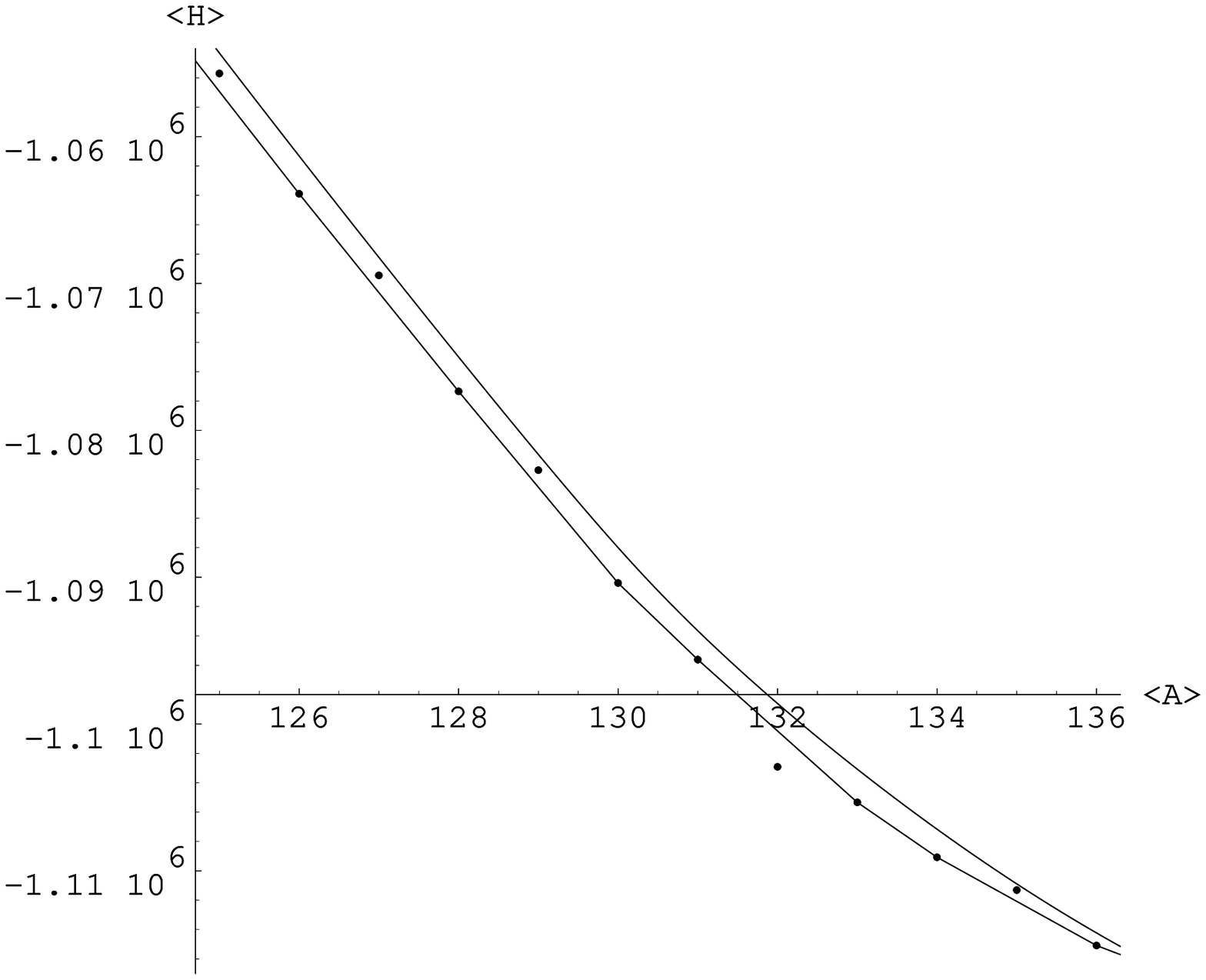}
     }
\vspace{-2cm}
\caption{Function $\langle {\bf H} \rangle(\langle {\bf A} \rangle)$ from
bare data. Left, including $^{132}$Sn. Right, without $^{132}$Sn. Lower
curves, $T=20$ keV. Upper ones, $T=2$ MeV.  Note the failure of the low
temperature curves at reproducing ground-state energies of odd nuclei.}
\end{figure}

\vspace{-3.5cm}
\begin{figure}[htb] \centering
\mbox{  \epsfysize=120mm
         \epsffile{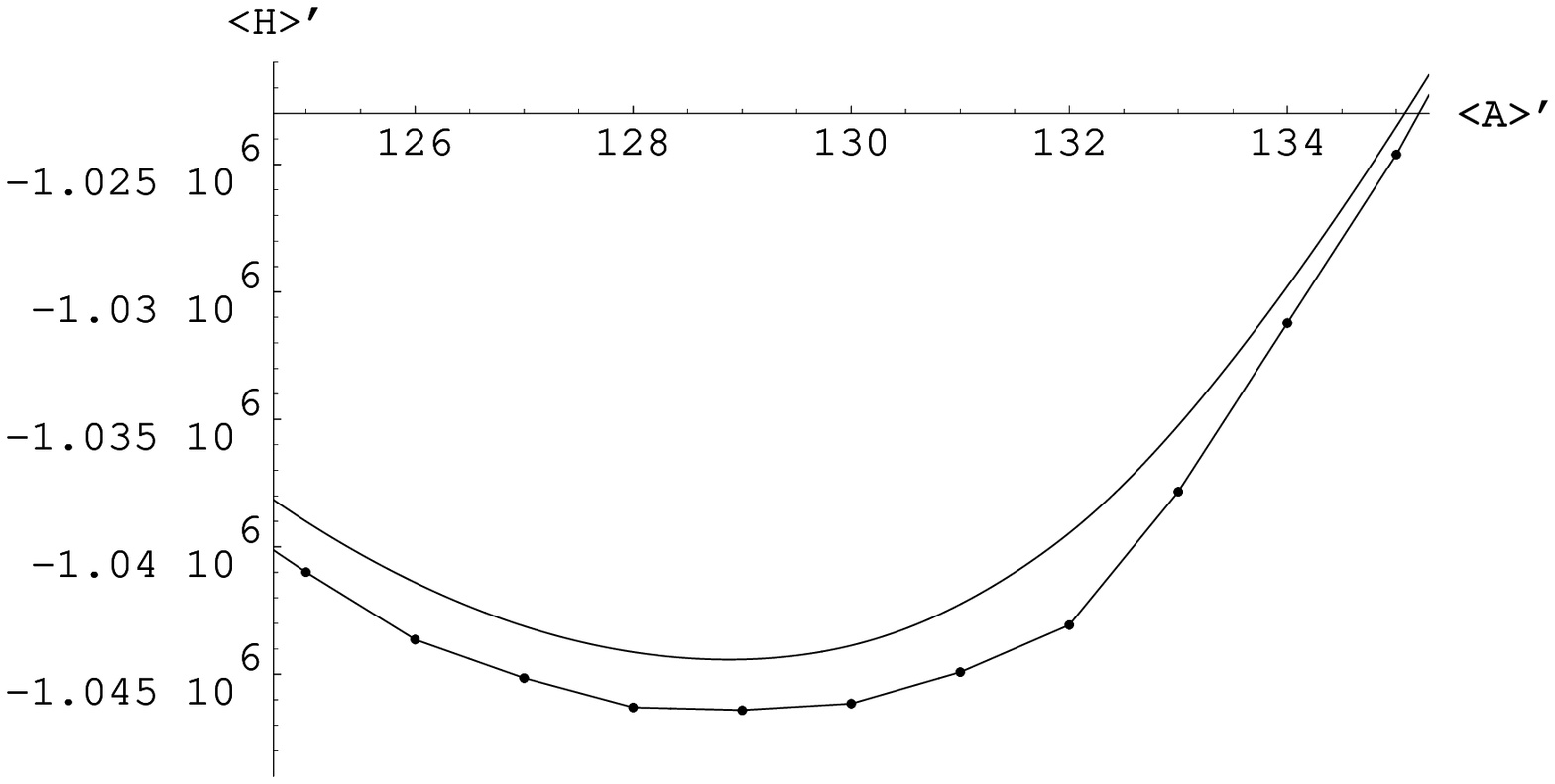}
     }
\mbox{  \epsfysize=120mm
         \epsffile{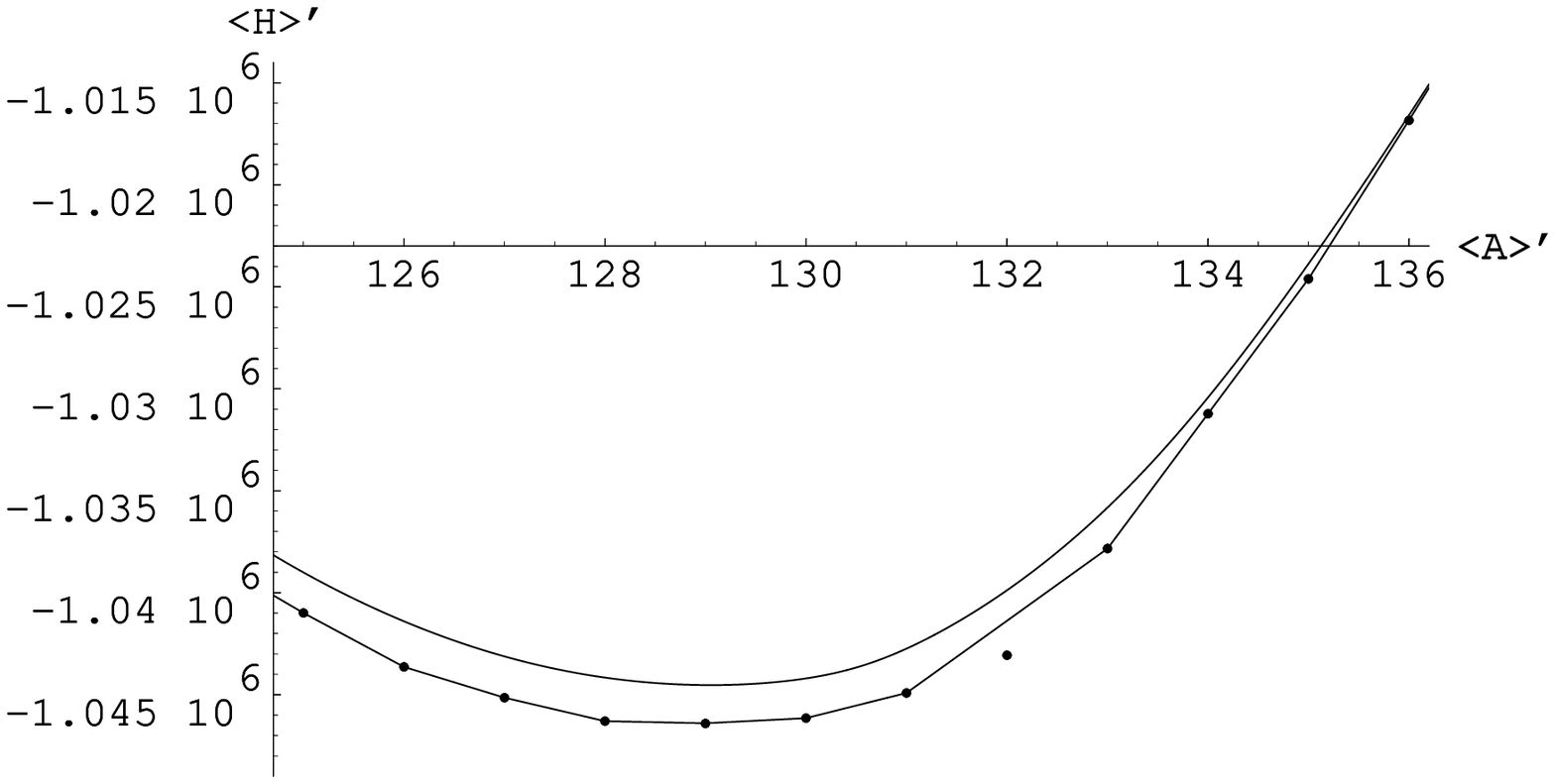}
     }
\vspace{-3.5cm}
\caption{Concave data $\langle {\bf H} \rangle'(\langle {\bf A} \rangle').$ 
Left, with $^{132}$Sn. Right, without $^{132}$Sn. Lower curves, $T=20$ keV.
Upper ones, $T=2$ MeV. In contrast with Fig. 1, the lower curves reproduce
the  ground-state energies of both odd and even contributor nuclei.}
\end{figure}

\vspace{-3cm}
\begin{figure}[htb] \centering
\mbox{  \epsfysize=120mm
         \epsffile{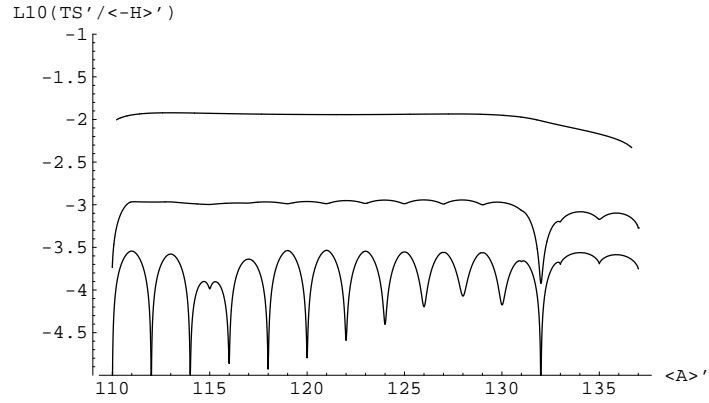}
     }
\vspace{-3cm}
\caption{Decimal log plots of 
$|T\, S'/\langle {\bf H} \rangle'| (\langle {\bf A} \rangle');$ upper curve,
$T=2$ MeV; intermediate one, $T=300$ keV; lower one, $100$ keV.}
\end{figure}

\vspace{-3cm}
\begin{figure}[htb] \centering
\mbox{  \epsfysize=120mm
         \epsffile{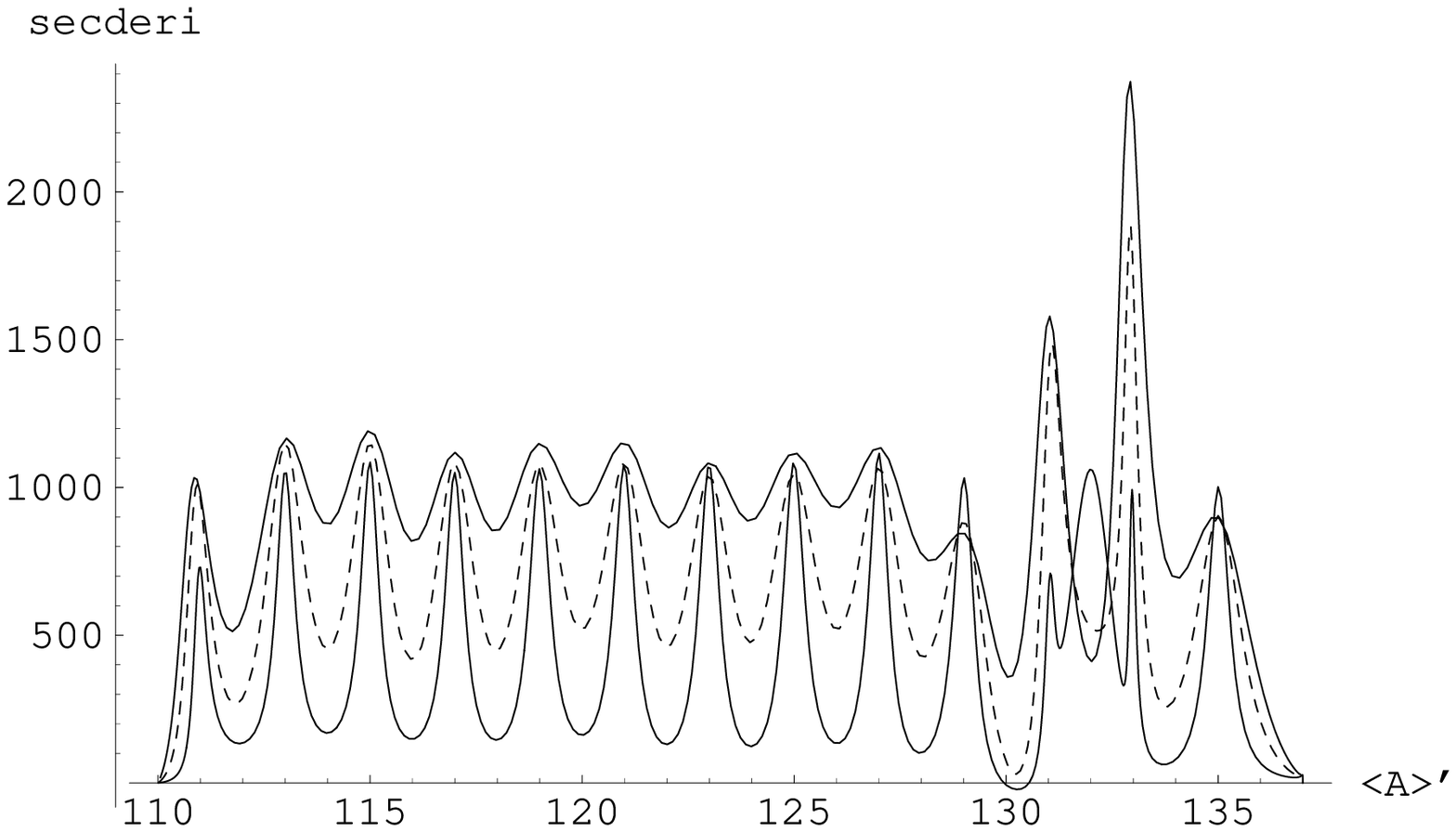}
     }
\mbox{  \epsfysize=120mm
         \epsffile{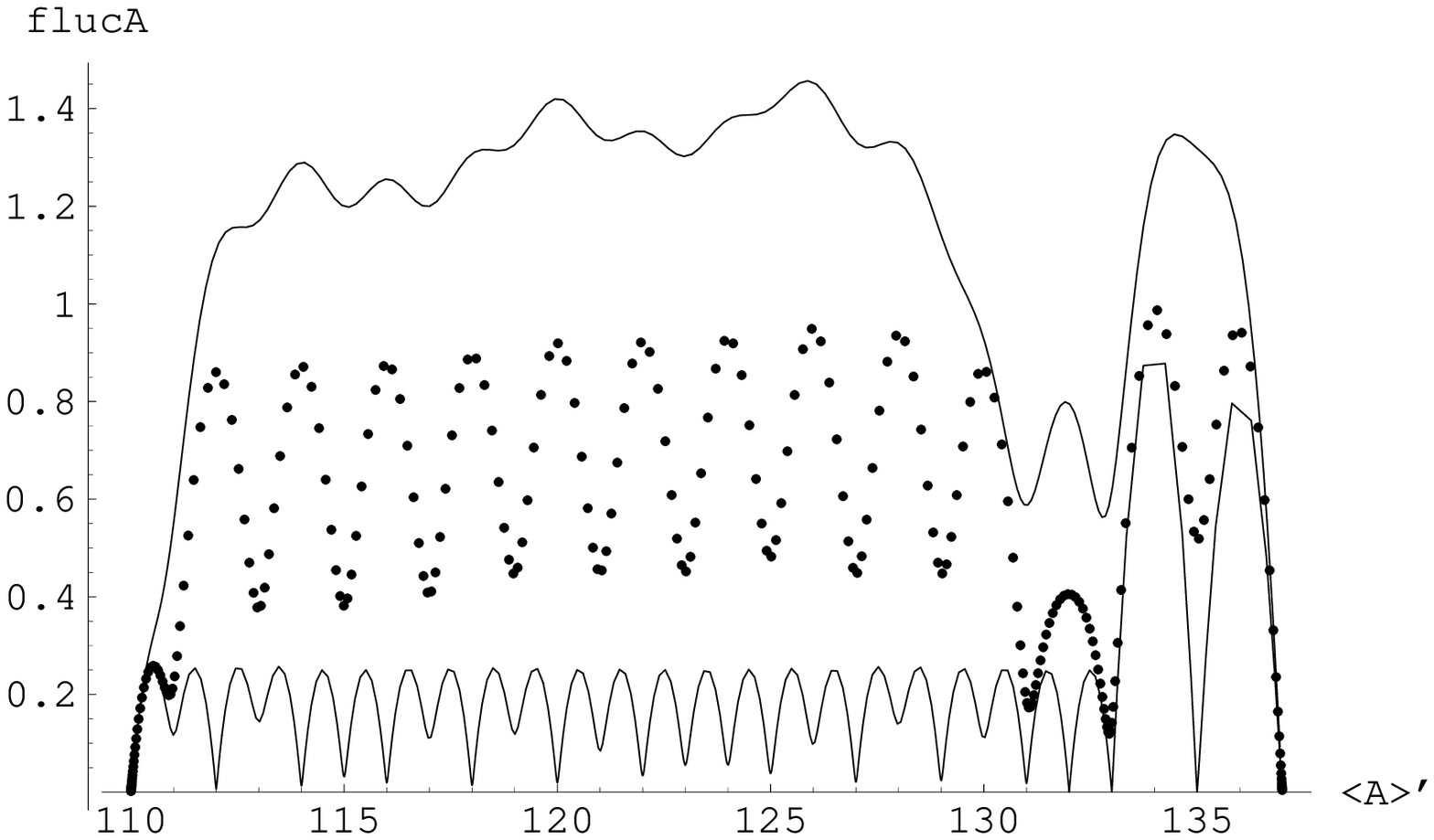}
     }
\vspace{-3cm}
\caption{Left, behavior of 
$\left( \langle {\bf A}^2 \rangle' - \langle {\bf A} \rangle'^2 \right)
\partial^2 \langle {\bf H} \rangle' / \partial \langle {\bf A} \rangle'^2;$
upper full line, $T=1$ MeV; dashed curve $T=750$ keV; lower full line,
$500$ keV. Right, behavior of
$\langle {\bf A}^2 \rangle' - \langle {\bf A} \rangle'^2$; upper curve,
$T=1$ MeV; dots, $T=500$ keV; lower curve, $100$ keV.}
\end{figure}

\vspace{-3cm}
\begin{figure}[htb] \centering
\mbox{  \epsfysize=120mm
         \epsffile{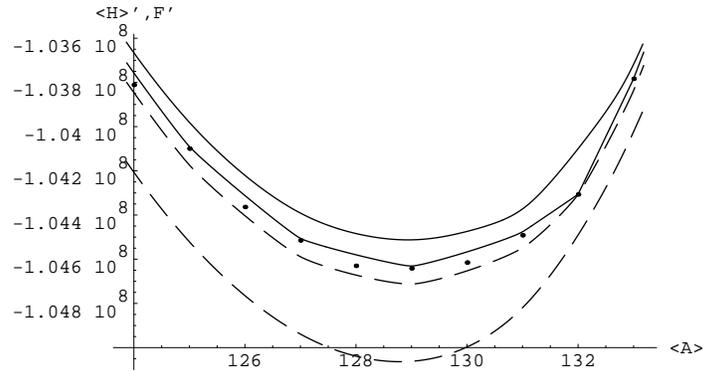}
     }
\vspace{-3cm}
\caption{Upper full curve, average energy $\langle {\bf H} \rangle'$ for
$T=1$ MeV, lower full one, same for $T=250$ keV. Upper dashed curve, free
energy ${\bf F}'$ for $T=250$ keV, lower dashed one, same for $T=1$ MeV.}
\end{figure}

\vspace{-3cm}
\begin{figure}[htb] \centering
\mbox{  \epsfysize=120mm
         \epsffile{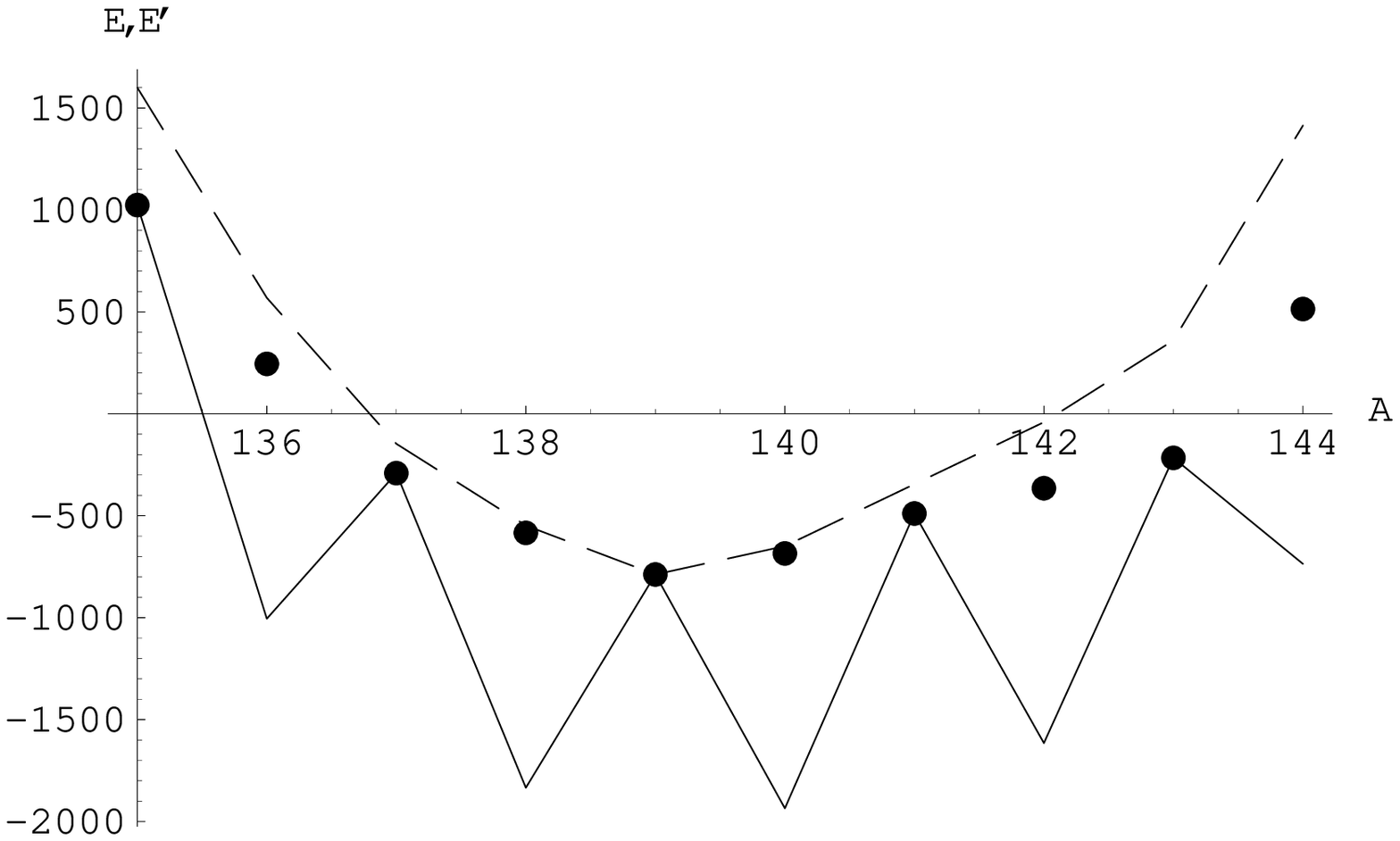}
     }
\mbox{  \epsfysize=120mm
         \epsffile{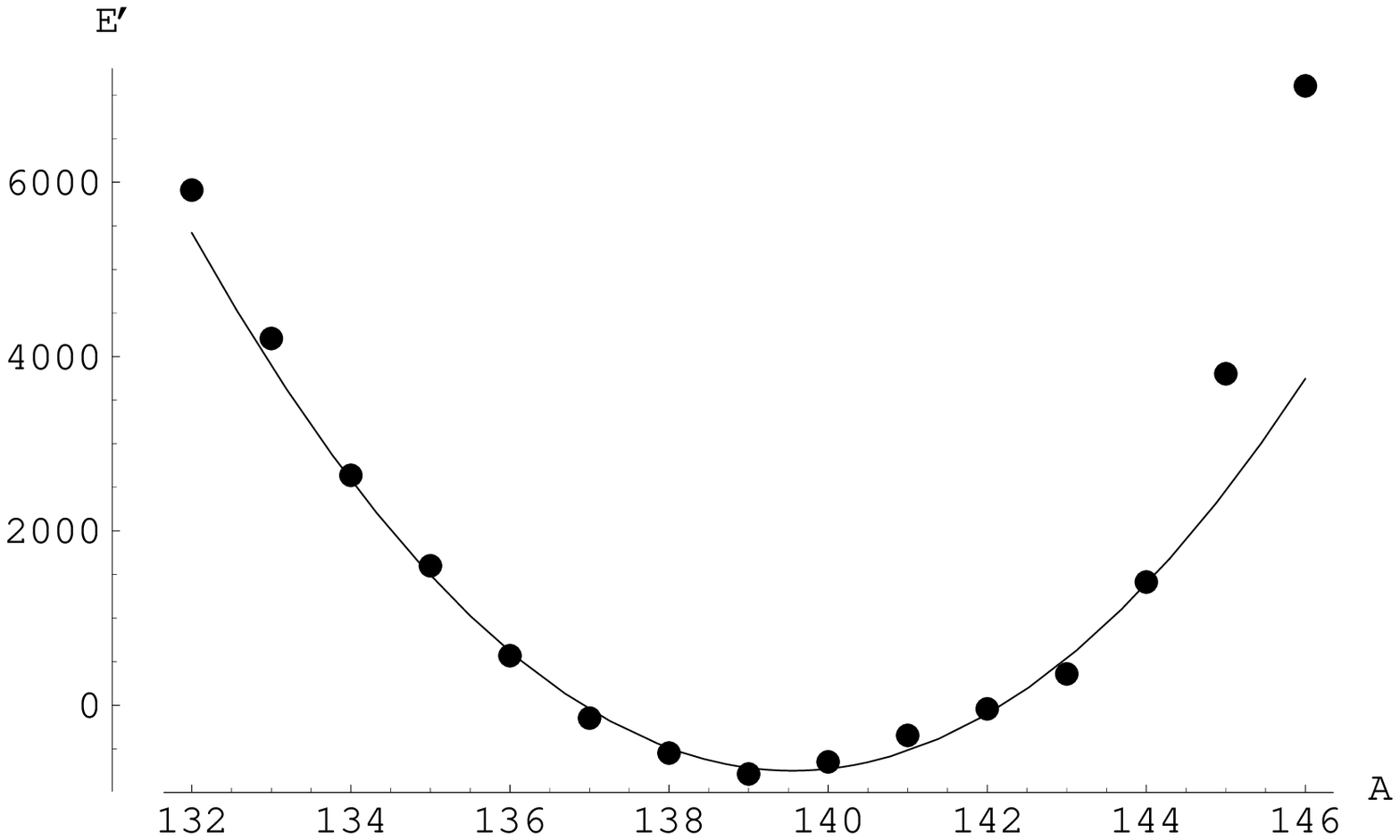}
     }
\vspace{-3cm}
\caption{Left: the full line joins Sm energies $E_A + 10000 \times A - 245000;$
the dots show the result of the pairing correction; the long dashed line 
connects tuned, ``concave energies'' $E^{\, \prime}_A.$ Right: dots, same 
tuned energies, from $^{132}$Sm to $^{146}$Sm; full line, least square fit by 
degree 2 polynomial between $^{135}$Sm and $^{144}$Sm only.}
\end{figure}

\vspace{-3cm}
\begin{figure}[htb] \centering
\mbox{  \epsfysize=120mm
         \epsffile{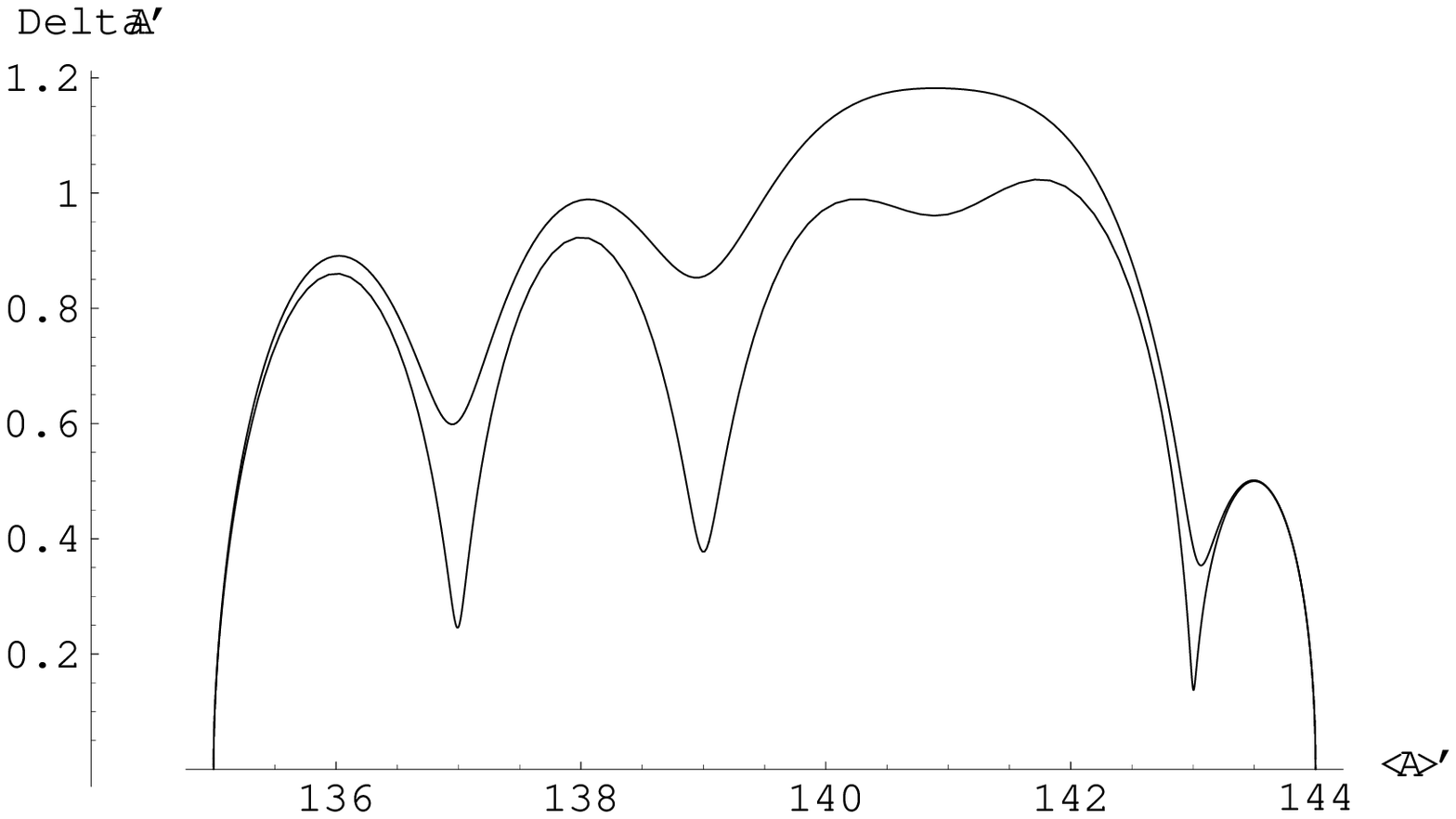}
     }
\mbox{  \epsfysize=120mm
         \epsffile{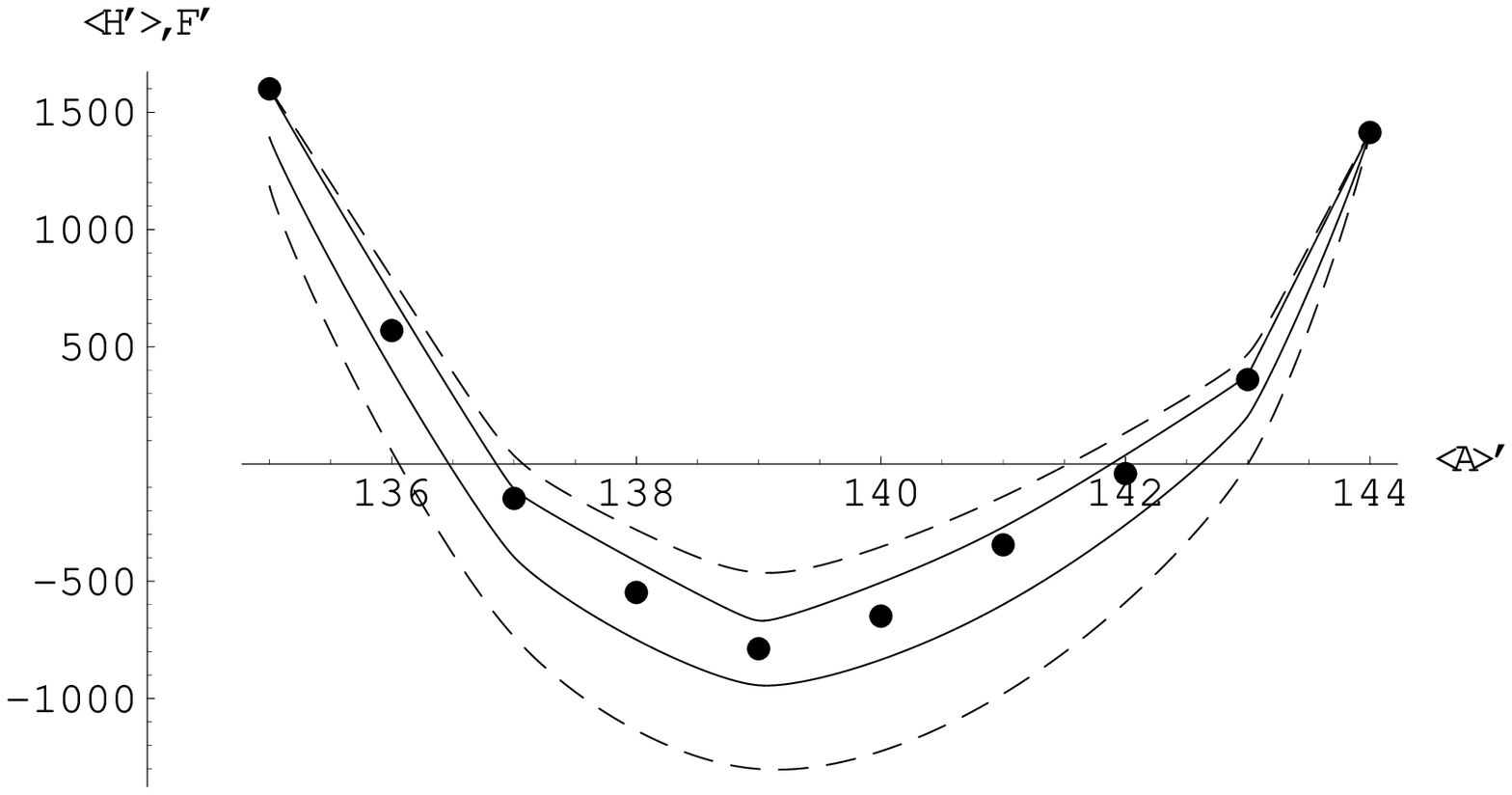}
     }
\vspace{-3cm}
\caption{Left: Sm particle number fluctuation at $200$ (upper curve) and
$100$ keV (lower curve). Right: dots, Sm ground-state tuned energies; upper
two curves, $\langle{\bf H} \rangle';$ lower two curves, ${\bf F}';$ dashed 
curves, $T=200$ keV, full ones, $T=100$ keV.}
\end{figure}

\vspace{-3cm}
\begin{figure}[htb] \centering
\mbox{  \epsfysize=120mm
         \epsffile{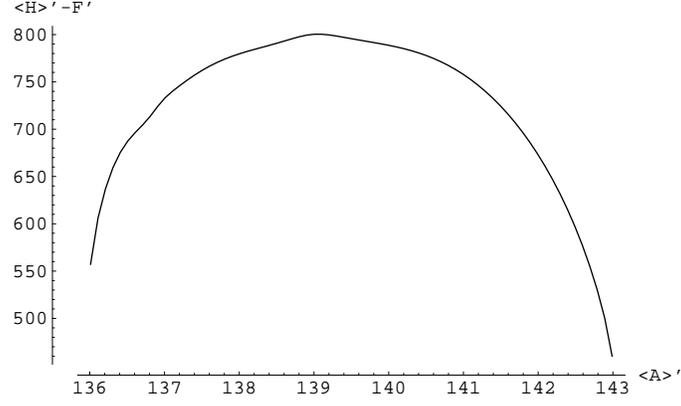}
     }
\vspace{-3cm}
\caption{Estimation of an error bar from the difference,
$\langle {\bf H} \rangle'_{300}-{\bf F}'_{150},$ as a function of 
$\langle {\bf A} \rangle'.$}
\end{figure}

\vspace{-3cm}
\begin{figure}[htb] \centering
\mbox{  \epsfysize=120mm
         \epsffile{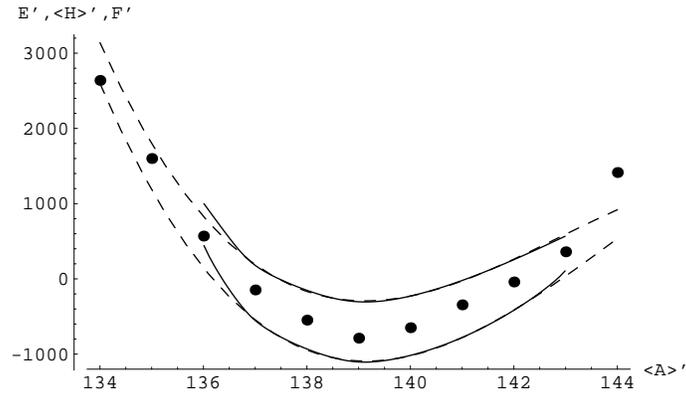}
     }
\vspace{-3cm}
\caption{Full curves: $\langle {\bf H} \rangle'_{300}$ (upper) and 
${\bf F}'_{150}$ (lower), calculated with eight spectra, from $^{136}$Sm to
$^{143}$Sm. Dashed curves: cubic polynomials obtained by least square fits 
between $^{137}$Sm and $^{142}$Sm. Dots: actual values $E'_A.$}
\end{figure}

\end{document}